
\magnification1200
\font\BBig=cmr10 scaled\magstep2
\font\BBBig=cmr10 scaled\magstep3


\def\title{
{\bf\BBBig
\centerline{Conformal Properties}\bigskip
\centerline{of}\bigskip
\centerline{Chern-Simons Vortices}\bigskip
\centerline{in}\bigskip
\centerline{External Fields}
}}

\def\foot#1{
\footnote{($^{\the\foo}$)}{#1}\advance\foo by 1
} 
\def\ccr{\cr\noalign{\medskip}}


\def\authors{
\centerline{
C. DUVAL\foot{D\'epartement de Physique, Universit\'e
d'Aix-Marseille II and
Centre de Physique
\hfill\break
Th\'eorique, CNRS-Luminy, Case 907, F-13288 MARSEILLE,
Cedex 09 (France)
\hfill\break
e-mail:duval@marcptsu1.univ-mrs.fr.}
P. A. HORV\'ATHY\foot{D\'epartement de Math\'ematiques,
Universit\'e de Tours, Parc de Grandmont,
\hfill\break
F-37200 TOURS (France) e-mail: horvathy@univ-tours.fr}
L. PALLA\foot{Institute for Theoretical Physics,
E\"otv\"os University, H-1088 BUDAPEST,
\hfill\break
Puskin u. 5-7 (Hungary).
e-mail: palla@ludens.elte.hu}}}

\def\runningauthors{Duval, Horv\'athy, Palla}
\def\runningtitle{Time-dependent Chern-Simons\dots}


\voffset = 1cm 
\baselineskip = 14pt 

\headline ={
\ifnum\pageno=1\hfill
\else\ifodd\pageno\hfil\tenit\runningtitle\hfil\tenrm\folio
\else\tenrm\folio\hfil\tenit\runningauthors\hfil
\fi
\fi}

\nopagenumbers
\footline={\hfil} 


\def\and{\qquad\hbox{and}\qquad}

\def\kikezd{\parag\underbar}

\def\IR{{\bf R}}

\def\smallcirc{{\raise 0.5pt \hbox{$\scriptstyle\circ$}}}
\def\smallover#1/#2{\hbox{$\textstyle{#1\over#2}$}}
\def\2{{\smallover1/2}}
\def\ccr{\cr\noalign{\medskip}}
\def\parag{\hfil\break}
\def\semidirectproduct{
{\ooalign{\hfil\raise.07ex\hbox{s}\hfil\crcr\mathhexbox20D}}
}
\def\={\!=\!}
\def\boxit#1{
\vbox{\hrule\hbox{\vrule\kern3pt
\vbox{\kern3pt#1\kern3pt}\kern3pt\vrule}\hrule}
} 

\def\cJ{{\cal J}}
\def\cB{{\cal G}}
\def\cG{{\cal G}}
\def\cP{{\cal P}}
\def\cD{{\cal D}}
\def\cK{{\cal K}}
\def\cH{{\cal H}}
\def\cM{{\cal N}}


\newcount\ch 
\newcount\eq 
\newcount\foo 
\newcount\ref 

\def\chapter#1{
\parag\eq = 1\advance\ch by 1{\bf\the\ch.\enskip#1}
}

\def\equation{
\leqno(\the\ch.\the\eq)\global\advance\eq by 1
}

\def\reference{
\parag [\number\ref]\ \advance\ref by 1
}

\ch = 0 
\foo = 1 
\ref = 1 


\title
\vskip 1.5cm
\authors
\vskip .25in

\parag
{\bf Abstract.}\hskip 2mm
{\it The construction and the symmetries of Chern-Simons vortices
in harmonic and uniform magnetic force backgrounds
found by Ezawa, Hotta and Iwazaki, and by Jackiw and Pi
are generalized using the non-relativistic Kaluza-Klein-type framework
presented in our previous paper. All Schr\"odinger-symmetric backgrounds
are determined.}

\vskip.3in


\vskip.3in

\centerline{\it Submitted to Physics Letters {\bf B}}

\vfill\eject


\chapter{Introduction}

The construction of static, non-relativistic Chern-Simons solitons [1]
was recently  generalized to time-dependent solutions,
yielding vortices in a constant external magnetic field,
${\cal B}$ [2-4]. Putting $\omega={\cal B}/2$,
the equation to be solved is
$$
i\big({D_\omega}\big)_t\Psi_\omega=\left\{
-{1\over2}{\vec{D}}_\omega^2
-\Lambda\,\Psi_\omega^*\Psi_\omega
\right\}\Psi_\omega.
\equation
$$
(We use units where $e=m=1$). Here the covariant derivative means
$$
\big({D_\omega}\big)_\alpha
\=
\partial_\alpha-i({A_\omega})_\alpha-i{\cal A}_\alpha
\equation
$$
($\alpha=0,1,2$), where ${\cal A}_\alpha$ is a vector
potential for the constant magnetic field,
${\cal A}_0=0$, ${\cal A}_i=
\2\epsilon_{ij}x^j{\cal B}\equiv\omega\epsilon_{ij}x^j$
($i,j=1,2$)
and
$(A_\omega)_\alpha$ is the vector potential
of Chern-Simons electromagnetism i.e. its field
strength is required to satisfy the field-current identity
$$
B_\omega\equiv
\epsilon^{ij}\partial_iA_\omega^j
=-{1\over\kappa}\rho_\omega
\and
E_\omega^i\equiv
-\partial_iA_\omega^0-\partial_tA_\omega^i
={1\over\kappa}\epsilon^{ij}J_\omega^j
\equation
$$
with
$\rho_\omega\=\Psi_\omega^*\Psi_\omega$
and
$\vec{J}_\omega\=
({1/2i})[\Psi^*\vec{D}_\omega
\Psi_\omega-\Psi_\omega(\vec{D}_\omega\Psi_\omega)^*]
$.
These equations can be solved [2-4] by applying a
coordinate transformation to a solution, $\Psi$, of the problem
with $\omega=0$ studied in Ref. [1], according to
$$\eqalign{
\Psi_\omega(t,\vec{x})&={1\over\cos\omega t}\,
\exp\left\{-i\omega{r^2\over2}\tan{\omega t}\right\}\,
\exp\left\{i{{\cal N}\over2\pi\kappa}\omega t\right\}\,
\Psi(\vec{X},T),
\ccr
(A_\omega)_\alpha&=A_\beta{\partial X^\beta\over\partial x^\alpha}
-\partial_\alpha\Big({\omega\over2\pi\kappa}{\cal N}t\Big),
\cr}
\equation
$$
with
$$
T={\tan\omega t\over\omega},
\qquad
\vec{X}={1\over\cos\omega t}R(\omega t)\,\vec{x}.
\equation
$$
Here ${\cal N}=\int\Psi^*\Psi\,d^2\vec{x}$ is the vortex number
and $R(\theta)$ is the matrix of a rotation by angle $\theta$ in the
plane.
(The pre-factor $\exp[i{\cal N}\omega t/2\pi\kappa]$ and the extra term
$-\partial_\alpha[(\omega/2\pi\kappa){\cal N}t]$ are absent
from the corresponding formula of Ezawa et al. [2]).
A similar construction works in a harmonic background [4].

In a previous paper [5]
non-relativistic Chern-Simons theory
in $2+1$ dimensions was obtained by reduction from an appropriate
$(3+1)$-dimensional Lorentz manifold. As an application, we
reproduced the results in Ref. [1].
Here we show that the above generalizations
arise by reduction from suitable curved spaces, that
they all share the (extended) Schr\"odinger
symmetry of the model in [1], and we determine
all background fields which have this property.

\goodbreak


\chapter{Chern-Simons theory in Bargmann space}

$(2+1)$-dimensional non-relativistic Chern-Simons theory can be lifted to
\lq Bargmann space' i.e. to a $4$-dimensional Lorentz manifold $(M,g)$
endowed with a
covariantly constant null vector $\xi$ [5].
The theory is described by a massless non-linear wave equation
$$
\Big\{D_\mu D^\mu
-{R\over6}
+\lambda\psi^*\psi\Big\}\psi=0,
\equation
$$
where $D_\mu=\nabla_\mu-ia_\mu$ ($\mu=0,1,2,3$),
$\nabla$ is the metric-covariant derivative and
$R$ denotes the scalar curvature.
The scalar field $\psi$ and the
`electromagnetic' field strength,
$f_{\mu\nu}=2\partial_{[\mu}a_{\nu]}$,
are related by the identity
$$
\kappa f_{\mu\nu}=
\sqrt{-g}\,\epsilon_{\mu\nu\rho\sigma}\xi^\rho j^\sigma,
\qquad
j^\mu={1\over2i}\left[\psi^*(D^\mu\psi)-\psi(D^\mu\psi)^*\right].
\equation
$$
Eq.~(2.1) [but {\it not} (2.2)] can be obtained
from variation of the `partial' action
$$
S={1\over2}\int_M\left\{(D_\mu\psi)^*\,D^\mu\psi
+{R\over 6}\,|\psi|^2-{\lambda\over 2}\,|\psi|^4
\right\}\sqrt{-g}\,d^4\!x.
\equation
$$

The quotient of $M$ by the integral curves of
$\xi$ is non-relativistic space-time we denote by $Q$.
A Bargmann space admits local coordinates $(t,\vec{x},s)$
such that $(t,\vec{x})$ label $Q$ and
$\xi=\partial_s$.
When supplemented by the equivariance condition
$$
\xi^\mu D_\mu\psi=i\psi,
\equation
$$
our theory projects to a non-relativistic
non-linear Schr\"odinger/Chern-Simons
theory on the $(2+1)$-dimensional manifold $Q$.
The field strength $f_{\mu\nu}$ is clearly the lift of a
closed two-form $F_{\mu\nu}$ on $Q$. So, the vector potential may be chosen as
$a_\mu=(A_\alpha,0)$ with $A_\alpha$ $s$-independent. In this gauge,
$\Psi(t,\vec{x})=e^{-is}\psi(t,\vec{x})$ is then a function on $Q$.

A {\it symmetry} is a transformation of $M$
which interchanges the solutions of the coupled system. Each
$\xi$-preserving conformal transformation is a symmetry and the
variational derivative
$\vartheta_{\mu\nu}=2\delta S/\delta g^{\mu\nu}$ provides us with a
conserved,
traceless and symmetric energy-momentum tensor.
The version of
Noether's theorem proved in [5] says that for any $\xi$-preserving
conformal vectorfield
$(X^\mu)$ on Bargmann space, the quantity
$$
Q_X=\int_{\Sigma_t}{
\vartheta_{\mu\nu}X^\mu\xi^\nu\,\sqrt{\gamma}\,d^2\vec{x}
}
\equation
$$
(where the `transverse space' $\Sigma_t$ is a space-like 2-surface
$t={\rm const.}$ and
$\gamma_{ij}$ is the metric induced on it by $g_{\mu\nu}$) is a constant
of the motion. The conserved quantities are conveniently calculated
using the formula [5]
$$
\vartheta_{\mu\nu}\xi^\nu
={1\over2i}\left[
\psi^*\,(D_\mu\psi)-\psi\,(D_\mu\psi)^*\right]
-{1\over 6}\,\xi_\mu\left(
{R\over 6}|\psi|^2
+(D^\nu\psi)^*\,D_\nu\psi
+{\lambda\over2}\,|\psi|^4
\right).
\equation
$$

For example, $M$ can be flat Minkowski space with metric
$d\vec{X}^2+2dTdS$, where
$\vec{X}\in\IR^2$ and $S$ and $T$ are light-cone coordinates.
This is the
Bargmann space of a free, non-relativistic particle [6].
The system of equations (2.1,2,4) projects in this case to that of
Ref. [1]; the
$\xi$-preserving conformal transformations form the (extended) planar
Schr\"o\-din\-ger group, consisting of the Galilei group with generators
$\cJ$ (rotation),
$\cH$ (time translation),
$\vec{\cG}$ (boosts),
$\vec{\cP}$ (space translations),
augmented with dilatation, $\cD$, and expansion, $\cK$, and
centrally extended by `vertical' translation, $\cM$. With a
slight abuse of notation, the associated conserved quantities are
denoted by the same symbols.
(Explicit formul{\ae} are listed in [1] and [5]).
Applying any symmetry transformation to a solution of the field equations
yields another solution. For example, a boost or an expansion applied
to the static solution $\Psi_0(\vec{X})$ of Jackiw and Pi
produces time-dependent solutions. Using the formul{\ae}
in [6] we find
$$
\Psi(T,\vec{X})={1\over1-kT}\exp\left\{-{i\over2}\Big[
2\vec{X}\cdot\vec{b}+T\vec{b}^2+k{(\vec{X}+\vec{b}T)^2\over1-kT}
\Big]\right\}\,\Psi_0({\vec{X}+\vec{b}T\over1-kT}),
\equation
$$
which is the same as in [1].

Now we present some new results.
The most general `Bargmann' manifold was found long
ago by Brinkmann [7]:
$$
g_{ij}dx^idx^j+2dt\big[ds+\vec{\cal A}\cdot d\vec{x}\big]
+2{\cal A}_0dt^2,
\qquad
{\cal A}_0=-U
\equation
$$
where the `transverse' metric $g_{ij}$ as well as the `vector potential'
$\vec{\cal A}$ and the `scalar' potential $U$
are functions of $t$ and $\vec{x}$ only.
Clearly, $\xi=\partial_s$ is a covariantly constant null vector.
The null geodesics of this metric describe particle motion
in curved transverse space in an external electromagnetic fields
$\vec{\cal E}=-\partial_t\vec{\cal A}-\vec{\nabla}U$
and ${\cal B}=\vec{\nabla}\times\vec{\cal A}$ [6].

Consider now a Chern-Simons vector potential
$(a_\omega)_\mu=\big((A_\omega)_\alpha,0\big)$
in the background (2.8).
[The subscript $(\,\cdot\,)_\omega$ refers to an
external-field problem].
Using that the only non-vanishing components of the metric (2.8) are
$g^{ij},\, g^{is}=-A^i$,
$g^{ss}=2U+A_iA^i$,
$g^{st}=1$,
we find that the integrand in the partial action (2.3) is
$$
\big(\vec{D}_\omega\psi\big)^*\cdot\vec{D}_\omega\psi
+i\Big[\big((D_\omega)_t\psi\big)^*\psi
-\psi^*(D_\omega)_t\psi\Big]
+{R\over6}\,|\psi|^2-{\lambda\over 2}\,|\psi|^4,
\equation
$$
where the covariant derivative $(D_\omega)_\alpha$ means
(1.2) with vector
potential ${\cal A}_\alpha$. Thus, including the `vector-potential'
components into the metric (2.8) results, after reduction, simply
in modifying the covariant derivative $D_\alpha$ in `empty' space
(${\cal A}_\alpha=0$) according to
$
D_\alpha\to(D_\omega)_\alpha.
$
The associated equation of motion is hence Eq.~(1.1).
(This conclusion can also be reached
directly by studying the wave equation (2.1)).

Let now $\varphi$ denote a conformal Bargmann
diffeomorphism between two
Bargmann spaces i.e. let $\varphi:(M,g,\xi)\to(M',g',\xi')$ be
such that $\varphi^\star g'=\Omega^2g$ and $\xi'=\varphi_\star\xi$.
Such a mapping projects to a diffeomorphism of the quotients,
$Q$ and $Q'$ we denote by $\Phi$.
Then the same proof as in Ref. [5] allows one to show that if
$(a'_\mu,\psi')$ is a solution of the field equations on
$M'$, then
$$
a_\mu=(\varphi^\star a')_\mu
\and
\psi=\Omega\,\varphi^\star\psi'
\equation
$$
is a solution of the analogous equations on $M$.
Locally we have
$\varphi(t,\vec{x},s)=(t',\vec{x}',s')
\equiv\big(\Theta(t),\Phi(t,\vec{x}),s+\Sigma(t,\vec{x})\big)$
so that
$\psi=\Omega\,\varphi^\star\psi'$ reduces to
$$
\Psi(t,\vec{x})=\Omega(t)\,e^{i\Sigma(t,\vec{x})}\Psi'(t',\vec{x}'),
\qquad
A_\alpha=\Phi^\star A'_\alpha
\qquad(\alpha=0,1,2).
\equation
$$
Note that $\varphi$ takes a $\xi$-preserving conformal
transformation of $(M,g,\xi)$ into a $\xi'$-preserving conformal
transformation of $(M',g',\xi')$.
Conformally related Bargmann spaces
have therefore isomorphic symmetry groups.

The  associated conserved quantities can be related by comparing
the expressions in Eq.~(2.6). Note first that, for $\psi$ as in Eq.~(2.10),
$
D_\mu\psi=\Omega\,(\varphi^\star D'_\mu\psi')+
\Omega^{-1}\nabla'_\nu\Omega\,\varphi^\star\psi'
$.
Using
$
R=\Omega^2\,\varphi^\star R'+6\Omega^{-1}\nabla'_\nu\nabla'{}^\nu\Omega
$
and
$
\xi_\mu=\Omega^{-2}\xi'_\mu
$
as well as $\Omega=\Omega(t)$ and
that $g^{\mu t}$ is non-vanishing only for $\mu=s$
one finds hence that
$
\vartheta_{\mu\nu}\xi^\nu=\Omega^2\varphi^\star
\big(\vartheta'_{\mu\nu}\xi'{}^\nu\big).
$
But
$
\sqrt{\gamma}=\Omega^{-2}\varphi^\star\sqrt{\gamma'}
$.
Therefore, the conserved
quantity (2.5)
associated to $X=(X^\mu)$ on $(M,g,\xi)$ and to $X'=\varphi_\star X$ on
$(M',g',\xi')$ coincide,
$$
Q_X=\varphi^\star Q'_{X'}.
\equation
$$
The labels of the generators
are, however, different (see the examples below).

\goodbreak


\chapter{Flat examples}

Consider now the Lorentz metric
$$
d\vec{x}_{osc}^2+2dt_{\rm osc}ds_{\rm osc}-\omega^2r_{\rm osc}^2dt_{\rm osc}^2
\equation
$$
where $\vec{x}_{\rm osc}\in\IR^2$, $r_{\rm osc}=|\vec{x}_{\rm osc}|$
and $\omega$ is a constant. Its null geodesics
correspond to a non-relativistic,
spinless particle in an oscillator background [6,9].
Requiring equivariance (2.4), the wave equation (2.1) reduces to
$$
i\partial_{t_{\rm osc}}\Psi_{\rm osc}=\left\{
-{\vec{D}^2\over2}+{\omega^2\over2}{r_{\rm osc}}^2
-\Lambda\,\Psi_{\rm osc}\Psi_{\rm osc}^*\right\}\Psi_{\rm osc}
\equation
$$
($\vec{D}=\vec{\partial}-i\vec{A}$, $\Lambda=\lambda/2$),
which describes Chern-Simons vortices in a harmonic force background,
studied in Ref. [3].
The clue is that the mapping
$\varphi(t_{\rm osc},\vec{x}_{\rm osc},s_{\rm osc})
=(T,\vec{X},S)$ [9], where
$$
T={\tan\omega\,t_{\rm osc}\over\omega},
\qquad
\vec{X}={\vec{x}_{\rm osc}\over\cos\omega t_{\rm osc}},
\qquad
S=s_{\rm osc}-{\omega r_{\rm osc}^2\over2}\tan\omega t_{\rm osc}
\equation
$$
carries the oscillator metric (3.1) conformally into the free form,
$
d\vec{X}^2+2dTdS
$,
with conformal factor
$\Omega(t_{\rm osc})=|\cos\omega t_{\rm osc}|^{-1}$
such that $\varphi_\star\partial_{s_{\rm osc}}=\partial_S$.
Our formula lifts the
coordinate transformation of Ref. [4] to Bargmann space.

A solution in the harmonic background can be
obtained by Eq.~(2.11).
A subtlety arises, though:
the mapping (3.3) is many-to-one:
it maps each `open strip'
$$
I_j=\big\{
(\vec{x}_{\rm osc},t_{\rm osc},s_{\rm osc})\,\big|
\,(j-\2)\pi<\omega t_{\rm osc}<(j+\2)\pi
\big\},
\equation
$$
where $\,j=0,\pm1,\ldots$ corresponding to a half oscillator-period onto
the full Minkowski space.
Application of (2.11) with $\Psi$ an `empty-space' solution
yields, in each $I_j$, a solution, $\Psi^{(j)}_{\rm osc}$. However,
at the contact points
$t_j\equiv(j+1/2)(\pi/\omega)$, these fields may not match.
For example, for the `empty-space' solution obtained by an expansion,
Eq.~(2.7) with $\vec{b}=0,\,k\neq0$,
$$
\lim_{t_{\rm osc}\to t_j-0}\Psi^{(j)}_{\rm osc}=
(-1)^{j+1}{\omega\over k}
e^{-i{\omega^2\over2k}r_{\rm osc}^2}\Psi_0(-{\omega\over k}\vec{x})=
-\lim_{t_{\rm osc}\to t_j+0}\Psi^{(j+1)}_{\rm osc}.
\equation
$$
Then continuity is restored by including the `Maslov'
phase correction [10]:
$$
\left\{\eqalign{
\Psi_{\rm osc}(t_{\rm osc},\vec{x}_{\rm osc})&=
(-1)^{j}\,\displaystyle{1\over\cos\omega t_{\rm osc}}\,
\exp\left\{-{i\omega\over2}r_{\rm osc}^2\tan{\omega t_{\rm osc}}\right\}\,
\Psi(T,\vec{X})
\ccr
(A_{\rm osc})_0(t_{\rm osc},\vec{x}_{\rm osc})
&={1\over\cos^2\omega t_{\rm osc}}
\big[
A_0(T,\vec{X})-\omega\sin\omega t_{\rm osc}\;
\vec{x}_{\rm osc}\cdot\vec{A}(T,\vec{X})
\big],
\ccr
\vec{A}_{\rm osc}(t_{\rm osc},\vec{x}_{\rm osc})
&={1\over\cos\omega t_{\rm osc}}\,
\vec{A}(T,\vec{X}),
\cr}\right.
\equation
$$
where $j$ is as in (3.4).
Eq.~(3.6)
extends the result in Ref. [4] from $|t_{\rm osc}|<\pi/2\omega$
to any $t_{\rm osc}$. For the static solution in [1] or for that obtained
from it by
a boost, $\lim_{t_{\rm osc}\to t_j}\Psi^{(j)}_{\rm osc}=0$
and the inclusion of the correction factor is not mandatory.

Chern-Simons theory in the oscillator-metric
has again a Schr\"odinger symmetry,
whose generators are related to those in `empty' space as
$$
\left\{
\eqalign{
J_{\rm osc}&=\cJ,
\cr
H_{\rm osc}&={\cH}+\omega^2\cK,
\cr
(C_{\rm osc})_\pm&=
\left(\cH-\omega^2\cK\pm 2i\omega\cD\right),
\cr
(\vec{P}_{\rm osc})_\pm&=
\left(\vec{\cP}\pm i\omega\vec{\cB}\right),
\cr
N_{\rm osc}&=\cM.
\cr
}\right.
\equation
$$
The oscillator-Hamiltonian, $H_{\rm osc}$, is hence a combination of the
Hamiltonian and of the expansion valid for $\omega=0$, etc.
The generators $H_{\rm osc}$ and $(C_{\rm osc})_\pm$ span ${\rm o}(2,1)$ and
the $(\vec{P}_{\rm osc})_\pm$ generate the two-dimensional Heisenberg algebra
[9].
Eq.~(3.7) adds $(C_{\rm osc})_\pm$ and $(\vec{P}_{\rm osc})_\pm$
to the $J_{\rm osc}$ and $H_{\rm osc}$ in Ref.~[3].

Consider next the metric
$$
d\vec{x}{}^2+2dt\Big[ds+
\2\epsilon_{ij}{\cal B}{x}^jd{x}^i\Big]
\equation
$$
where $\vec{x}\in\IR^2$ and ${\cal B}$
is a constant. Its null geodesics
describe a charged particle in a uniform magnetic
field in the plane [6].
Again, when
imposing equivariance, Eq.~(2.1)
reduces precisely
to Eq.~(1.1) with $\Lambda=\lambda/2$ and
covariant derivative $D_\omega$ given as in Eq.~(1.2).
The metric (3.8) is readily transformed into an
oscillator metric (3.1):
the mapping
$\varphi(t,\vec{x},s)=(t_{\rm osc},\vec{x}_{\rm osc},s_{\rm osc})$
given by
$$
t_{\rm osc}=t,
\qquad
x_{\rm osc}^i=x^i\cos\omega t+\epsilon^i_jx^j\sin\omega t,
\qquad
s_{\rm osc}=s
\equation
$$
--- which amounts to switching
to a rotating frame with angular velocity $\omega={\cal B}/2$ ---
takes the \lq constant ${\cal B}$-metric' (3.8) into the
oscillator metric (3.1).
The vertical vectors
$\partial_{s_{\rm osc}}$ and $\partial_s$ are permuted.
Thus, the time-dependent rotation (3.9)
followed by the transformation (3.3),
which projects to the coordinate
transformation (1.5) of Refs. [2] and [3],
carries conformally the
constant-${\cal B}$ metric (3.8) into the $\omega\=0$-metric.
It
carries therefore
the \lq empty' space so\-lution $e^{is}\Psi$ with $\Psi$ as
in (2.7) into that in
a uni\-form mag\-netic field back\-ground
according to Eq.~(2.10). Taking into account the equivariance,
we get the formul{\ae} of [2] i.e. (1.4) without the
${\cal N}$-terms --- but multiplied with the Maslov factor
$(-1)^j$. (The
${\cal N}$-term arises due to a subsequent
gauge transformation required by the gauge fixing in [3]).

It also allows to
\lq export' the Schr\"odinger symmetry to
non-relativistic Chern-Simons theory in
the constant magnetic field background. The [rather
complicated] generators, listed in Ref. [11], are readily obtained
using Eq.~(2.12). For example, time-translation $t\to t+\tau$ in
the ${\cal B}$-background amounts to a time translation for the
oscillator with parameter $\tau$ plus a rotation with angle $\omega\tau$.
Hence
$
H_{\cal B}=H_{\rm osc}-\omega\cJ=\cH+\omega^2\cK-\omega\cJ.
$
Similarly, a space translation for ${\cal B}$ amounts, in `empty' space,
to a space translations and a boost, followed by a rotation, yielding
$P_B^i=\cP^i+\omega\,\epsilon^{ij}\cG^j$, etc.

\goodbreak


\chapter{Conformally flat Bargmann spaces}

All our preceding results apply to any Bargmann space which can
be conformally mapped into Minkowski space in a $\xi$-preserving way.
Now we describe these
`Schr\"odinger-conformally flat' spaces.
In $D=n+2>3$ dimensions, conformal
flatness is guaranteed by the
vanishing of the conformal Weyl tensor
$$
C^{\mu\nu}_{\ \ \rho\sigma}
=
R^{\mu\nu}_{\ \ \rho\sigma}
-
\smallover 4/{D-2}\,\delta^{[\mu}_{\ [\rho}\,R^{\nu]}_{\ \sigma]}
+
\smallover 2/{(D-1)(D-2)}\,
\delta^{[\mu}_{\ [\rho}\,\delta^{\nu]}_{\ \sigma]}\,R.
\equation
$$
Now $R_{\mu\nu\rho\sigma}\xi^\mu\equiv 0$ for a Bargmann space, which
implies some extra conditions
on the curvature. Inserting the identity
$\xi_\mu R^{\mu\nu}_{\ \ \rho\sigma}=0$
into
$C^{\mu\nu}_{\ \ \rho\sigma}=0$,
using the identity $\xi_\mu R^\mu_\nu\equiv 0$
($R^\nu_\sigma\equiv R^{\mu\nu}_{\ \ \mu\sigma}$),
we find
$
0=
-\left[
\xi_\rho R^\nu_\sigma-\xi_\sigma R^\nu_\rho\right]
+
R/(D-1)\left[
\xi_\rho\delta^\nu_\sigma-\xi_\sigma\delta^\nu_\rho
\right]
$.
Contracting again with $\xi^\sigma$ and using that $\xi$ is null,
we end up with
$R\,\xi_\rho\xi^\nu=0$. Hence the scalar curvature vanishes, $R=0$.
Then the previous equation yields
$\xi^{ }_{[\rho}R^\nu_{\sigma]}=0$ and thus
$R_\sigma^\nu=\xi_\sigma\eta^\nu$
for some vector field $\eta$. Using the
symmetry of the Ricci tensor, $R_{[\mu\nu]}=0$,
we find that $\eta=\varrho\,\xi$
for some function $\varrho$. We finally get the {\it consistency relation}
$$
R_{\mu\nu}=\varrho\,\xi_\mu\xi_\nu.
\equation
$$
The Bianchi identities ($\nabla_\mu R^\mu_\nu=0$
since $R=0$) yield $\xi^\mu\partial_\mu\varrho=0$, i.e. $\varrho$ is
a function on spacetime $Q$.
The conformal Schr\"odinger-Weyl tensor is hence of the
form
$$
C^{\mu\nu}_{\ \ \rho\sigma}=
R^{\mu\nu}_{\ \ \rho\sigma}
-
\smallover 4/{D-2}\,
\varrho\,\delta^{[\mu}_{\ [\rho}\,\xi^{\nu]}\xi^{ }_{\sigma]}.
\equation
$$
It is noteworthy that Eq.~(4.2) is
the Newton-Cartan field equation with
$\varrho/(4\pi G)$ as the matter density of the sources.

It follows from Eq.~(4.2) that the transverse Ricci tensor
of a  Schr\"odinger-conformal flat Bargmann metric necessarily vanishes,
$R_{ij}=0$ for each~$t$.
The transverse space is hence (locally) flat and we
can choose $g_{ij}=g_{ij}(t)$.
Then a change of coordinates
$(\vec{x},t,s)\to(G(t)^{1/2}\vec{x},t,s)$ where $G(t)^i_j=g_{ij}(t)$
[which brings in a uniform magnetic field and/or an oscillator to the
metric],
casts our Bargmann metric into the form (2.8) with $g_{ij}=\delta_{ij}$
while $\xi$ remains unchanged.

The non-zero components of the Weyl tensor
of the general $D=4$ Brinkmann metric (2.8) are found as
$$
\eqalign{
&C_{xyxt}=-C_{ytts}=-\smallover1/4\partial_x{\cal B},
\ccr
&C_{xyyt}=+C_{xtts}=-\smallover1/4\partial_y{\cal B},
\ccr
&C_{xtxt}=-\2\big[
\partial_t(\partial_y{\cal A}_y-\partial_x{\cal A}_x)-
{\cal A}_x\partial_y{\cal B},
\big]
+\2\big[\partial_x^2-\partial_y^2\big]U,
\ccr
&C_{ytyt}=+\2\big[
\partial_t(\partial_y{\cal A}_y-\partial_x{\cal A}_x)
-{\cal A}_y\partial_x{\cal B}\big]
-\2\big[\partial_x^2-\partial_y^2\big]U,\ccr
&C_{xtyt}=+\2\big[
\partial_t(\partial_x{\cal A}_y+\partial_y{\cal A}_x)
+2\partial_x\partial_yU\big]
-\smallover1/4({\cal A}_x\partial_x-{\cal A}_y\partial_y){\cal B}.
\cr
}
\equation
$$
Then Schr\"odinger-conformal flatness requires
$$
\left\{\eqalign{
&{\cal A}_i=\2\epsilon_{ij}{\cal B}(t)x^j+a_i,
\qquad
\vec{\nabla}\times\vec{a}=0,
\qquad\partial_t\vec{a}=0,
\ccr
&U(t,\vec{x})=\2 C(t)r^2+\vec{F}(t)\cdot\vec{x}+K(t). \cr}
\right.
\equation
$$
(Note, {\sl en passant}, that (4.2) automatically holds:
the only non-vanishing component of the Ricci tensor is
$R_{tt}=
-\partial_t(\vec{\nabla}\cdot\vec{\cal A})-\2{\cal B}^2-\Delta U$).

This Schr\"odinger-conformally flat  metric hence allows one to describe a
uniform magnetic field ${\cal  B}(t)$, an attractive [$C(t)=\omega^2(t)$] or
repulsive [$C(t)=-\omega^2(t)$]
isotropic oscillator and a uniform force field $\vec{F}(t)$ in the plane
which may all depend arbitrarily on time. It also includes a curlfree
vector potential~$\vec{a}(\vec{x})$ that can be gauged away if the
transverse space is simply connected: $a_i=\partial_if$ and the coordinate
transformation $(t,\vec{x},s)\to(t,\vec{x},s+f)$
results in the `gauge'
transformation ${\cal{A}}_i\to{\cal{A}}_i-\partial_if=-\2{\cal
B}\,\epsilon_{ij}x^j$. However, if space is not simply connected, we can
also include an external Aharonov-Bohm-type vector potential.

Being conformally related, all these metrics share the symmetries of flat
Bargmann space: for example, if the transverse space is $\IR^2$ we get the
full Schr\"odinger symmetry; for $\IR^2\setminus\{0\}$ the symmetry is
reduced rather to
${\rm o}(2)\times{\rm o}(2,1)\times\IR$,
just like for a magnetic vortex [12].

The case of a constant electric field
is quite amusing. Its metric,
$d\vec{x}^2+2dtds-2\vec{F}\cdot\vec{x}dt^2$, can be brought to the free
form by switching to an accelerated coordinate system,
$$
\vec{X}=\vec{x}+\2\vec{F}\,t^2,
\qquad
T=t,
\qquad
S=s-\vec{F}\cdot\vec{x}\,t-\smallover1/6\vec{F}^2t^3.
\equation
$$
This example (chosen by Einstein to illustrate the equivalence principle)
also shows that the action of the Schr\"odinger group --- e.g. a rotation
--- looks quite different in the inertial and in the moving frames.

Let us finally mention that the above results admit a `gauge theoretic'
interpretation. In conformal (Lorentz) geometry, the Weyl tensor
$C_{\mu\nu\rho\sigma}$ arises as part of the
${\rm o}(n+2,2)$-valued curvature
of a Cartan connection for a $D=n+2$ dimensional base manifold. The
Schr\"odinger-conformal geometry in this dimension can be viewed as a reduction
of the standard conformal geometry to the Schr\"odinger subgroup
${\rm Sch}(n+1,1)\subset{\rm O}(n+2,2)$
[13]. The curvature of the reduced Cartan
connection then defines the Schr\"odinger-Weyl tensor which is thus
characterized by
$$
C_{\mu\nu\rho\sigma}\,\xi^\mu=0,
\equation
$$
a property coming from the previous embedding and strictly equivalent
to Eqs.~(4.2,3).

\chapter{Conclusion}

Our `non-relativistic Kaluza-Klein' approach provides a unified
view on the various vortex constructions in external fields,
explains the common origin of the large symmetries,
and allows us to describe all such spaces.
We have also pointed out that the formula (1.4) may require a slight
modification for times larger then a half oscillator-period.

\kikezd{Acknowledgement}. We are indebted to Professor R. Jackiw for
stimulating discussions.
L. P. would like to thank Tours University for the
hospitality extended to him
and to the Hungarian National Science and Research Foundation
(Grant No. $2177$) for a partial financial support.

\vfill\eject
\goodbreak
\vskip5mm
\centerline{\bf\BBig References}

\reference
R. Jackiw and S-Y. Pi,
Phys. Rev. Lett. {\bf 64}, 2969 (1990);
Phys. Rev. {\bf 42}, 3500 (1990).

\reference
Z. F. Ezawa, M. Hotta and A. Iwazaki,
Phys. Rev. Lett. {\bf 67}, 411 (1991);
Phys. Rev. {\bf D44}, 452 (1991).

\reference
R. Jackiw and S-Y. Pi,
Phys. Rev. Lett. {\bf 67}, 415 (1991).

\reference
R. Jackiw and S-Y. Pi, Phys. Rev. {\bf 44}, 2524 (1991).

\reference
C. Duval, P. A. Horv\'athy and L. Palla,
Phys. Lett. {\bf B325}, 39 (1994).

\reference
C. Duval, G. Burdet, H-P. K\"unzle and M. Perrin,
Phys. Rev. {\bf D31}, 1841 (1985); see also
C. Duval, G. Gibbons and P. Horv\'athy,
Phys. Rev. {\bf D43}, 3907 (1991).

\reference
H.W. Brinkmann, Math. Ann. {\bf 94}, 119 (1925).

\reference
R. Jackiw, Phys. Today {\bf 25}, 23 (1972);
U. Niederer, Helv. Phys. Acta {\bf 45}, 802 (1972);
C. R. Hagen, Phys. Rev. {\bf D5}, 377 (1972);
G. Burdet and M. Perrin, Lett. Nuovo Cim. {\bf 4}, 651 (1972).

\reference
U. Niederer, Helv. Phys. Acta {\bf 46}, 192 (1973);
G. Burdet, C. Duval and M. Perrin,
Lett. Math. Phys. {\bf 10}, 255 (1986);
J. Beckers, D. Dehin and V. Hussin, J. Phys. {\bf A20}, 1137 (1987).

\reference
V. P. Maslov, {\it The Theory of Perturbations and Asymptotic Methods}
(in Russian), izd. MGU: Moscow, (1965);
J-M Souriau, in Proc {\it Group Theoretical Methods in Physics},
Nijmegen'75, Janner (ed) Springer Lecture Notes in Physics {\bf 50}, (1976).

\reference
M. Hotta, Prog. Theor. Phys. {\bf 86}, 1289 (1991).

\reference
R. Jackiw, Ann. Phys. (N. Y.) {\bf 201}, 83 (1990).

\reference
M. Perrin, G. Burdet and C. Duval,
Class. Quant. Grav. {\bf 3}, 461 (1986);
C. Duval, in {\it Conformal Groups and Related Symmetries. Physical
Results and Mathematical Background}, ed.
A. O. Barut and H. D. Doebner, Lecture Notes in Physics {\bf 261}, p. 162
Berlin: Springer-Verlag (1986).

\vfill\eject

\bye